# Probabilistic Photonic Computing with Chaotic Light

*Frank Brückerhoff-Plückelmann[1], Hendrik Borras[2], Bernhard Klein[2], Akhil Varri[1], Marlon Becker[3,4], Jelle Dijkstra[5], Martin Brückerhoff[6], C. David Wright[7], Martin Salinga[8], Harish Bhaskaran[9], Benjamin Risse[3,4], Holger Fröning[2], Wolfram Pernice[1,5]\**

[1]Physical Institute, University of Münster; Münster, 48149, Germany.
[2]Institute of Computer Engineering, University of Heidelberg; Heidelberg, 69120, Germany.
[3]Institute for Geoinformatics, University of Münster; Münster, 48149, Germany.
[4]Faculty of Mathematics & Computer Science, University of Münster; Münster, 48149, Germany
[5]Kirchhoff-Institute for Physics, University of Heidelberg; Heidelberg, 69120, Germany.
[6]DEVK RE; Cologne, 50668, Germany.
[7]Department of Engineering, University of Exeter; Exeter, EX44QF, UK.
[8]Institute of Materials Physics, University of Münster; Münster, 48149, Germany.
[9]Department of Materials, University of Oxford; Oxford, OX43PJ, UK.

\*Correspondence to: wolfram.pernice@kip.uni-heidelberg.de

**Biological neural networks effortlessly tackle complex computational problems and excel at predicting outcomes from noisy, incomplete data, a task that poses significant challenges to traditional processors. Artificial neural networks (ANNs), inspired by these biological counterparts, have emerged as powerful tools for deciphering intricate data patterns and making predictions. However, conventional ANNs can be viewed as "point estimates" that do not capture the uncertainty of prediction, which is an inherently probabilistic process. In contrast, treating an ANN as a probabilistic model derived via Bayesian inference poses significant challenges for conventional deterministic computing architectures. Here, we use chaotic light in combination with incoherent photonic data processing to enable high-speed probabilistic computation and uncertainty quantification. Since both the chaotic light source and the photonic crossbar support multiple independent computational wavelength channels, we sample from the output distributions in parallel at a sampling rate of 70.4 GS/s, limited only by the electronic interface. We exploit the photonic probabilistic architecture to simultaneously perform image classification and uncertainty prediction via a Bayesian neural network. Our prototype demonstrates the seamless cointegration of a physical entropy source and a computational architecture that enables ultrafast probabilistic computation by parallel sampling.**

## Introduction

According to the neuroscience principle of free energy minimization (FEM), living organisms develop internal models of their environment to guide actions that minimize surprise and reduce uncertainty [1,2]. This objective stands in contrast to that of biologically inspired artificial neural networks (ANNs), which typically aim to maximize accuracy [3]. Shifting focus from accuracy to handling uncertainty is pivotal in explaining the efficiency and adaptability of biological neural networks. To date, ANNs have been very successfully implemented on deterministic conventional hardware and have led to breakthrough results in areas including weather forecasting [4], medical diagnostic [5], autonomous driving [6] and natural language processing [7–10]. However, deterministic models are point estimates based on known data and do not take the complete posterior distribution of the parameters into account [11]. Bayesian neural networks (BNNs) replace the deterministic network parameters with probability distributions to capture the probabilistic nature of inferring from incomplete observed data [12,13]. In this way, BNNs allow for distinguishing between epistemic uncertainties due to the lack of data and aleatoric uncertainties arising from noise in the data itself [14,15]. Consequently, BNNs are also significantly more robust against overfitting to small data sets [16,17]. Bayesian inference also lies at the heart of the of the FEM principle.

Processing complex probabilistic models poses major challenges for conventional deterministic hardware. Because the integral formulations used in describing probabilistic models become intractable already for a small number of parameters, Monte Carlo methods are employed to provide approximate solutions [13,16]. This includes sampling from the model's posterior distribution multiple times and subsequently evaluating the model for each drawn sample. Thus, high-speed (true) random number generators are required in combination with an architecture capable of evaluating the full model for each sample in a reasonable time. In conventional hardware implementation, one major factor contributing to the inefficiency of machine learning systems is the reliance on the von Neumann digital architecture, which, contrary to the physics of computing substrates, enforces determinism and separates memory from computation [18]. Brain-inspired computing differs from conventional digital computing by emphasizing in-memory analog processing, fine-grained parallelism, reduced precision, increased randomness, adaptability, analog processing, and possibly, spike-based communication [19]. Co-designing FEM-based learning with brain-inspired computing platforms can enhance energy efficiency and adaptability by shifting the learning objective from noise reduction (accuracy) to instead harnessing hardware noise as a valuable computational resource [19]. For electronic crossbar arrays, memristors serve as the main in-memory computation element due their tunable conductance. Simultaneously, programming and reading the conductance of a memristor is a stochastic process due to inherent randomness of the switching process in addition to drifts and instabilities [20,21]. Since the randomness is programmable by deploying multiple memristors for a single matrix weight, it can be deployed for Bayesian inference. In this case, sampling from the posterior distribution is implemented by reading, and potentially rewriting, the memristor several times while the neuromorphic crossbar architecture ensures the efficient evaluation of the model [22]. To avoid the need for sequential sampling and the random structural changes within memristive materials, transitioning to the optical domain allows for probabilistic computing in parallel with single-shot readout by deploying chaotic light. Chaotic light is an ideal entropy source for true random number generation [23–26] and can, moreover, easily be generated at default telecom wavelengths by amplified spontaneous emission in erbium doped fibers or erbium doped waveguides [27–31]. Moreover, the incoherent nature and large optical bandwidth of chaotic light allows for high-speed data processing in photonic crossbar arrays by exploiting wavelength division multiplexing.

In the following we present a photonic neuromorphic architecture capable of performing probabilistic single-shot computations with a photonic crossbar array. We harness chaotic light fields as the entropy source of the system and as the optical carrier for probabilistic information encoding. For photonic in memory computing, we employ the non-volatile phase change material Germanium-Antimony-Telluride (GST). Using time-amplitude modulation, we perform probabilistic data encoding and achieve parallel sampling based on spectral demultiplexing. We quantify the precision of the stochastic multiply and accumulate operations performed by the photonic circuit. With an incoherent photonic processor, we calculate high-speed probabilistic convolutions on visual inputs, making use of parallel spectral sampling from the output distributions. We deploy stochastic variational inference in a Bayesian neural network based on the LeNet 5 [32] architecture to minimize the divergence between the true posterior of our model parameters and the variational distributions educible by our encoding scheme. We benchmark the BNN's accuracy and out-of-domain rejection on an incomplete MNIST [33] data set.

**System Architecture**

Photonic probabilistic computing relies on the capability to generate analog signals which encode input vectors with tailored mean and variance. In order to provide the desired mean/variance tuples we employ chaotic light as the entropy source of our system. In a chaotic light source, the beating between the various frequency components leads to a time varying optical intensity. Since the variance of the intensity fluctuations is proportional to the squared mean intensity, desired intensity distributions can be conveniently shaped by amplitude modulation. In addition, chaotic light offers the unique possibility to tune the autocorrelation of the fluctuations by varying the optical bandwidth. The correlation between samples drawn from the fluctuation approaches zero for sampling rates smaller than the optical bandwidth, since the coherence time, given by the Wiener-Khinchin theorem, is approximately the inverse bandwidth, see **Extended Data Fig. 3**. Making use of wavelength division multiplexing (WDM), we can employ a single chaotic light source, which easily spans an optical bandwidth of several THz, to provide multiple independent entropy sources. Since the computing mechanism is solely intensity-based, incoherent photonic crossbar arrays which are naturally broadband and compatible with broadband chaotic light can be used for efficient probabilistic computing. Photonic crossbars support parallel computing by WDM [34] and thus, a single chaotic light source can serve as the entropy source for all parallel computation channels.

**Fig. 1** sketches the working principle of the probabilistic processor. As a chaotic light source, we split the amplified spontaneous emission (ASE) of an Erbium doped fiber into four different waveguides and delay the signals with respect to each other beyond the coherence time with fiber loops. In this way, the superposition of the chaotic fields behaves like a single chaotic field with a mean intensity given by the sum of the individual intensities, see **Extended Data Fig. 4**. A desired input distribution is encoded in a sequence of pulse shapes which are modulated onto the optical carrier signals via electro-optic modulators (EOMs). These input distributions form the vector entries to the photonic crossbar which is used to perform the matrix vector multiplications (MVM). MVMs form the backbone of arithmetic operations in artificial neural networks and our photonic architecture is designed to do this in fast and efficient way. Within the photonic crossbar array, the matrix weights are encoded into an optical attenuation using appropriate states of the non-volatile phase change material Germanium-Antimony-Telluride (GST). Here, we perform additions (accumulations) by overlapping propagating pulses in a single waveguide and multiplications by attenuating the pulses with GST cells corresponding to matrix weights. At the output of the crossbar, we demultiplex the broadband ASE light according to the 200 GHz ITU grid. Since we perform the encoding on all frequency components in parallel at the input and only demultiplex the field before detection at the output, all wavelength channels carry the same intensity distribution. Therefore, WDM enables independent parallel sampling from the output distribution while minimizing data-shuffling.

**Probabilistic Encoding**

We drive the EOMs with a symbol rate of 17.6 GBaud and sample in parallel from the four ITU wavelength channels C28, C30, C32 and C34 with a readout circuit electronically limited to 30 GHz. **Fig. 2a** shows the mean and standard deviation of the photodetector signal for ten symbols encoded in subsequent time slots separated by 56.8 ps measured in C28. There are two main contributions to the measured output distribution for a given mean $\bar{x}$. First, there are intensity fluctuations due to the chaotic carrier signal which are described by a M-fold Bose-Einstein distribution. The fluctuations in the measured signal are proportional to the mean intensity value. The degeneracy factor M depends on the number of independent temporal coherence cells within a measurement interval and is therefore linked to the optical bandwidth and the electrical bandwidth of the readout circuit. Due to the high photon numbers used in our experiments, the optical shot noise is negligible. Second, there is electronic ground noise which can be described by a Gaussian distribution with standard deviation $\sigma_{el}$. In an idealized system, the measured output distribution $p(x, \bar{x})$ is the convolution of those two independent random processes, given as:

$$p(x, \bar{x}) = \int_0^\infty \left[ \frac{M^M}{\bar{x} \cdot \Gamma(M)} \cdot \left(\frac{v}{\bar{x}}\right)^{M-1} \cdot e^{-M \cdot v/\bar{x}} \right] \cdot \left[ \frac{1}{\sqrt{2 \cdot \pi \cdot \sigma_{el}^2}} \cdot e^{-0.5 \cdot (x-v)^2/\sigma_{el}^2} \right] dv \quad \text{Eq. 1}$$

Since the mean $\bar{x}$ depends on the encoded input symbol, i.e., the transmission through the EOM, the chaotic carrier enables shaping of the measured output distribution by changing the programmed waveform. To describe the measured distributions, we also take system imperfections into account, such as the limited extinction ratio of the modulator and detector saturation as described in the Supplement. **Fig. 2b** shows the output distributions measured for different encoded means at the photodetector, fitted to the complete physical model derived in the Supplement. For zero mean (i.e. maximum attenuation by the EOM), electrical noise is the dominant source of randomness, thus leading to a Gaussian shape of the output distribution. With increasing mean, intensity fluctuations described by a M-fold Bose-Einstein distribution become the major contribution to the shape of the distribution. For the largest measured mean, detector saturation reduces the width of the distribution as the maximal measured voltage is limited.

Next, we investigate the behavior of independent samples drawn from the four WDM wavelength channels shown in **Fig. 2c** when sampling is performed in parallel. **Fig. 2d** shows the measured standard deviation of the output distribution in each channel in dependence of the mean of the distribution. For each wavelength the standard deviation follows the model prediction. Small differences between the output distributions are caused by slightly different spectral shapes of the WDM channels and the fact that there are four different readout circuits with slightly different ground noise. Since the ASE can be modelled as the superposition of independent random emitters with fixed wavelength, there is no correlation between the intensity fluctuations in different wavelength channels for ideal demultiplexing. Practically, the measured correlation coefficients between different channels during parallel sampling is below $10^{-2}$ as shown **Fig. 2e**.

For a single symbol the mean of the distribution is directly connected to the variance by equation 1. In order to generate mean values with desired variance, we take the sum of 9 subsequent symbols which enables shaping the distribution of the measured sum. **Fig. 2f** shows the distribution for three different mean-variance tuples encoded in channel 34. If we encode the mean of the distribution only in a single symbol and set all other to zero (dark blue trace), the distribution of the sum behaves like that of a single symbol (standard deviation of 0.47). In contrast, spreading the same mean over all 9 symbols (light blue trace) leads to a distribution with the same mean but lower variance as the noise partially averages out (standard deviation of 0.29). In this way, we can tune the mean and variance of the output distribution independently. The main advantage of this encoding scheme is that the electronic readout circuit always performs identical operations, i.e., summing over 9 received symbols, and does not require any information about the noise distribution. The distribution is solely encoded in the waveform encoded on the chaotic optical carrier at the input and is propagated through the circuit. We note, that employing longer time sequences with more symbols provides a wider tuning range in the variance at the cost of longer integration time and an increased impact of the ground noise.

**Photonic in-memory computing**

We employ a photonic crossbar array for probabilistic computation. The architecture exploits photonic in-memory computing with waveguide coupled GST nanocells used as memory and multiplication units. We tune the optical attenuation of the phase change material by partially switching it between its barely absorptive amorphous and highly absorptive crystalline state, see **Extended Data Fig. 5**. Since both states are non-volatile, GST enables in-waveguide multiplication without requiring a constant power supply to hold the memory state. **Fig. 3a** shows the concept of photonic multiplication operations performed on time-varying input waveforms. The optical pulses corresponding to a desired input value propagate through the waveguide and couple evanescently to the GST cell. Through the interaction, the pulses are attenuated by an amount depending on the phase state of the GST and are used for further processing. Besides multiplication, we perform photonic additions by overlapping two pulse shapes in a single waveguide as sketched in **Fig. 3b**. As described in the Supplement, the sum of two chaotic light fields behaves like a single field with a mean corresponding to the sum of the input means. Since both multiplication and addition operations are linear and (optical) phase-insensitive, we parallelize them via wavelength division multiplexing.

We analyze the multiplication operation by optically writing a relative transmission coefficient of 0.6 into the GST nanocell. We exemplary choose an input distribution with a mean of 1 and sample from the output distribution in four WDM channels in parallel. **Fig. 3c** shows the measured input and output distribution together with the model prediction for the different wavelength channels. As expected, decreasing the transmission through the GST cell decreases the mean of the distribution accordingly. In contrast, the standard deviation is not linearly decreased because the electronic ground noise of the readout circuit is independent from the optical intensity. Next, we investigate the accumulation, i.e. addition, of two input distributions. **Fig. 3d** shows the measured mean of each output distribution in dependence on the sum of the input distributions' means. As expected, the means of the distributions add up and align with the model prediction with an average error of 0.4 % and a spread of 2.6 %. In addition, we compare the standard deviation of the output distribution with the standard deviation of the input distributions. Because the electronic noise is independent of the mean of the distribution, the standard deviations do not add up in the same way as the means. When comparing the width of the output distribution with the model prediction, we observe good alignment with the model within 0.6 %, with a spread of 1.2%.

**Probabilistic Convolutions**

Moving beyond individual operations on distributions, we fabricate the 4x4 photonic crossbar array shown in **Fig. 4a** which combines multiple multiplication and addition units. The high parallelism of the crossbar architecture is well suited for convolution processing due to the shared weights [35]. In **Fig. 4b** we illustrate the application of programmable probabilistic convolution operation for a visual (image) input. As in the deterministic (convolutional) counterpart [35], we encode the convolution kernel weights in the GST cells and encode the image pixel values in the mean input intensities. In addition, we choose selectively how to spread the mean over the 9 subsequent input symbols contributing to the output distribution. Here, we employ an encoding represented by a probabilistic mask. In the outer area of the picture, we encode the mean given by the pixel value in a single symbol and set the remaining symbols to zero (wide standard deviation of 0.45 for mean 1). In the inner region we spread the mean over all nine symbols (narrow standard deviation 0.28 for mean 1). In this way, the output distributions in the outer area exhibit larger noise levels in the convolution output compared to the inner area. It is important to note that the electronic readout circuit always performs the same operation and does not depend on the noise level. The noise level is solely encoded in the waveform modulated on the carrier signal.

We optically program the kernel weights for average pooling into the GST nanocells within the crossbar and calculate probabilistic convolution on the input image with a stride of two. We encode the input vectors on four chaotic light carrier signals and sample from the output distribution after the crossbar in parallel with four wavelength channels. We perform both the signal encoding and sampling with 17.6 GBaud per channel. **Fig 4. c-f** show a sample from the convolutions output distribution for each wavelength channel, color coded by the ITU WDM channel as in **Fig. 2c**. As determined by the probabilistic mask, the average pooled output is qualitative noisier in the outer region of the picture than in the inner region. We quantitatively compare the pooling of pixels with identical means in both regions. As expected, the standard deviation is smaller for the same mean in the inner region. For both cases the output distribution aligns well with the model prediction.

**Bayesian Inference**

Integrating the photonic probabilistic computing architecture in a neural network enables out-of-domain (OOD) detection via Bayesian inference. In this way the network does not only generate a prediction but also quantifies the similarity between the input and previously observed data, providing a measure for the confidence of the neural network. We design a modified LeNet-5 [32] deep neural network architecture, shown in **Fig. 5a**, and deploy the MNIST dataset for training and benchmarking. The dataset contains images of handwritten digits, which are to be categorized into ten classes, representing the numbers zero to nine. We create an OOD scenario by training the network only with nine classes, numbers zero to eight, and deploy the handwritten nines as out-of-domain data. While many options for training via Bayesian Inference exist, we choose to make use of the natural similarity between stochastic variational inference (SVI) and the photonic accelerator. Variational inference approximates the true posterior of the model parameters by utilizing simpler distributions $q(w)$, in this case the probabilistic properties of the chaotic light. During training SVI maximizes the so-called evidence lower bound (ELBO) using backpropagation. The ELBO consists of two terms, the first term represents data likelihood, and the second term is the Kullback-Leibler (KL) divergence between the approximated posterior q(w) and the prior p(w). It can be broadly written as [36]:

$$\text{ELBO}(q) = \mathbb{E}[\log p(D|w)] - \text{KL}(q(w)||p(w)) \qquad \text{Eq(2)}$$

To accelerate the off-chip training, we approximate the complex photonic distribution shown in Equation 1 by a Gaussian distribution during training.

**Fig. 5b** shows the classification performance, as well as the average OOD performance during training. We use the intuitive metric of accuracy on a test subset of known classes to monitor classification performance. To determine how well the BNN distinguishes OOD images from in-domain (ID) images, we compare Mutual Information (MI) on test images of known against unknown classes [37,38]. The network quickly learns to correctly classify known images, as the test accuracy rises to over 99%. While the difference in Mutual Information between known (low MI) and unknown (high MI) samples improves at the same time, it is much slower and converges to a fixed difference after about 100 epochs. This shows that by using SVI training the BNN has effectively learned to correctly classify images of known class, while being able to identify OOD images.

Finally, we transfer the learned parameters to the physical encoding scheme shown in **Fig. 2f**, the exact description of the implementation is shown in the Supplement. **Fig. 5c** shows the output distributions for an ID image, clearly assigning the highest output scores to the correct classes. In contrast, the output distributions overlap for the unknown number nine as shown in **Fig. 5d**. ID and OOD data can be distinguished on a per-image basis using Mutual Information as shown in **Fig. 5e**. Evaluating the network again on the test subset leads to similar results as for the Gaussian approximation during training. The accuracy slightly decreases from 99.41% to 99.37 % whereas the relative difference in average mutual information between OOD and ID data increases from to x23.24 to x25.60.

**Discussion and conclusion**

The probabilistic photonic processing architecture outlined above enables parallel sampling of distributions at high speed dictated by telecom frequencies. In contrast to electronic probabilistic processors which employ the switching dynamics of stochastic magnetic tunnel junctions, hafnium-oxide-based filamentary memristors or phase change materials as an entropy source, chaotic light sources provide physical entropy with very high bandwidth. Differing from optical entropy sources, such electronic probabilistic approaches are limited by their sequential sampling process and the material properties, i.e., large switching times in comparison to optical encoding and limited endurance. Our approach overcomes those limitations by using a chaotic light source in combination with a broadband incoherent photonic crossbar array, encoding the distribution in subsequent temporal bins and deploying broadband computation with spectral demultiplexing at the outputs for parallel sampling. With an electronic bandwidth of 30 GHz, a symbol rate of 17.6 GBaud per channel and sampling from 4 channels in parallel, the effective sample rate from a single matrix output distribution is 70.4 GS/s. In comparison, the sample rate, i.e., the inverse time for programming and reading the underlying material entropy source, ranges from 500 MS/s [20] to 1 MS/s [21], which implies a speedup of more than 2 orders of magnitude with photonic sampling. From a computing perspective, employing chaotic light as a carrier leads to interesting scaling properties. Since the architecture is based on the beating between the frequency components of chaotic light, we can use the same optical wavelength channel for all crossbar inputs, thus decoupling the number of inputs from the required optical bandwidth. Note that in our experiments the number of parallel samples was mainly limited by the number of input channel of the oscilloscope used during the measurements.

The physical entropy source is a natural fit for stochastic variational inference (SVI) as one of the major methods for Bayesian neural networks since SVI allows for operation on arbitrary parametrized probability distributions. This enables us to design an SVI representation of the photonic processor, including probabilistic and deterministic parameters, to train a Bayesian neural network that allows to reason about uncertainties. In contrast to standard BNN implementations based on probabilistic weights, we show that we can adhere to hardware properties by employing learnable probabilistic activations in our BNN architecture. We demonstrate the effectiveness of this approach on a BNN architecture trained for image classification with simultaneous OOD detection.

A key feature of our architecture is that both, the computation of multiply and accumulate operations with the photonic crossbar array and the tunable noise generation with a chaotic light source, are passive transmission measurements. Thus, limitations as limited endurance and low sampling rating arising from entropy sources based on material switching dynamics do not apply. Since a photonic crossbar array is functional over a range of several THz [34,35], chaotic light sources easily support dozens of THz and only a single wavelength channel is needed to draw a sample from the output distribution. Hence, the overall speed is solely limited by the electronic interface. Overall, our approach to probabilistic computing provides an effective method to remove the computational bottleneck of probabilistic modelling with conventional deterministic hardware.

## Methods

**Nanofabrication**

We create the photonic chip design with gdshelpers (39), a Python-based open-source design framework for integrated circuits. Our material stack consists of HSQ cladded stoichiometric LPCVD Si3N4 films (330 nm) atop SiO2 dielectric (3300 nm) with Silicon serving as the substrate material. The wafers are obtained from Rogue Valley Microdevices and are annealed prior to fabrication to improve the quality of the Si3N4 film. The fabrication process encompasses four stages. Initially, we deposit gold markers for aligning the various masks with respect to each other. In the second stage, we pattern the photonics, followed by sputtering the phase-change material Germanium-Antimony-Tellurium (GST-225). Finally, we clad the waveguides with HSQ. For exposing the various resists, we deploy the 100kV Raith EBPG 5150 electron beam lithography tool.

To create the gold markers, we initiate the process by spin-coating the positive photoresist polymethyl methacrylate (PMMA) from the AllResist AR-P 672 series. Following resist baking, we expose the marker regions. Subsequently, we develop the resist in a methyl isobutyl ketone (MIBK) and isopropyl alcohol (IPA) solution and evaporate a stack of chromium (5nm) / gold (80 nm) / chromium (5nm) through physical vapor deposition (PVD). The chromium layers at the bottom and top enhance adhesion and protect the gold surface. We liftoff the unexposed areas via sonication in acetone. Next, we pattern the photonic circuit into the negative resist AR-N 7520.12 (Allresist), which is spin-coated with a thickness of 350 nm. After development in a MF-319 (Microposit) solution, we etch the mask into the silicon nitride layer via reactive ion etching in a CHF3/O2 plasma (Oxford PlasmaPro RIE 80). Then we remove the mask with oxygen plasma. We fabricate the PMMA mask for GST deposition in the same way as the mask for gold evaporation. After development, we deposit 10 nm of GST-225 covered by 10 nm of Al2O3, which locally confines the GST during melt-quenching and furthermore protects it from oxidation, via sputter-deposition. Next, we liftoff the unexposed areas with acetone. Finally, we spincoat and expose 800 nm of the negative resist HSQ/FOX16 (Dow Corning) to clad the photonic circuit.

**Experimental setup**

We deploy an Agiltron ASES-1611A3113 as a chaotic light source and filter is to the relevant wavelength region, C28/C30/C32/C34 of the ITU grid, upon amplifying the light with a PriTel

FA-33-IO. Afterwards we split the light to 4 input channels and delay the channels with at least 1.25ns with respect to each other. Then we modulate the pulse shapes on the chaotic carrier signal with OptiLab IML-1550-40-PM-V electro optic modulators. The EOMs are controlled by a Keysight M9502A. For each pulse shape we optimize the coupling to the chip by adjusting the polarization. To measure the output of the system, we amplify the signal with a PriTel LNTFA-20-NMA before splitting it to the four wavelength channels. For detection we deploy Thorlabs RXM38AF detectors which are connected to a Keysight DSA-X 95004Q to measure the optical intensity. The overall bandwidth of the detection system is limited to 30 GHz by the oscilloscope. We use the python interface of the arbitrary waveform generator and the oscilloscope to control the complete system by the PC as shown in **Extended Data Fig. 1**.

**Phase Change Photonics**

The photonic crossbar arrays consist of multiple cells as shown in **Extended Data Fig. 5a**, each representing one matrix weight. The input light corresponding to the vector component is coupled by a directional coupler to a crossing with integrated Germanium-Antimony-Telluride (GST) on top, which serves a tunable, non-volatile attenuator. Afterwards, the light is coupled by a directional coupler to the output waveguide again. The transmission through the GST crossing strongly depends on the phase state of the GST, which is highly absorptive in its crystalline state but only barely absorbs in the amorphous one. We can trigger a phase transition of the GST and hence tune the matrix by sending high power optical pulses through the GST cell. **Extended Data Fig. 5b** shows a typical programming of the GST to different transmission levels relative to the crystalline one. In a closed-loop way, we measure the transmission through the cell and adjust the power of the 200ns write pulse to obtain the desired weight. With pulse powers between 4mW and 14 mW we can set the transmission with an error below 1%.

## Acknowledgements

We thank Jochen Stuhrmann, from Illustrato, and Jonas Schütte for their assistance with the illustrations. Also, we thank Ivonne Bente and Niklas Vollmar for supporting us during the fabrication process. The research is funded by:

- German Research Foundation under Germany´s Excellence Strategy EXC 2181/1—390900948 (the Heidelberg STRUCTURES Excellence Cluster), the Excellence Cluster

- 3D Matter Made to Order (EXC-2082/1—390761711) and CRC 1459 "Intelligent matter"
- European Union's Horizon 2020 research and innovation programme (grant no. 101017237, PHOENICS project) and the European Union's Innovation Council Pathfinder programme (grant no. 101046878, HYBRAIN project).
- COMET program within the K2 Center "Integrated Computational Material, Process and Product Engineering (IC-MPPE) (Project No 886385)
- Austrian Federal Ministries for Climate Action, Environment, Energy, Mobility, Innovation and Technology (BMK) and for Labour and Economy (BMAW), represented by the Austrian Research Promotion Agency (FFG), and the federal states of Styria, Upper Austria and Tyrol

## Author contributions

- Conceptualization: FB, WP, HB, CDW, HF
- Methodology: FP, HB, BK, AV, MB, JD, MB
- Investigation: FP, HB, BK, AV, ML, JD, MB
- Visualization: FB HB, BK, JD
- Funding acquisition: WP, CDW, MS, HB, BR, HF
- Project administration: WP, HF
- Supervision: WP, CDW, MS, HB, BR, HF
- Writing – original draft: FB, HB, WP
- Writing – review & editing: All authors

## Competing interests

The authors declare that they have no competing interests.

## Additional information

Supplementary materials are available for this paper.

Extended data figures are available for this paper.

Correspondence and requests for materials should be addressed to W.P.

## Data availability

All data are available in the main text, supplementary materials, or extended data figures.

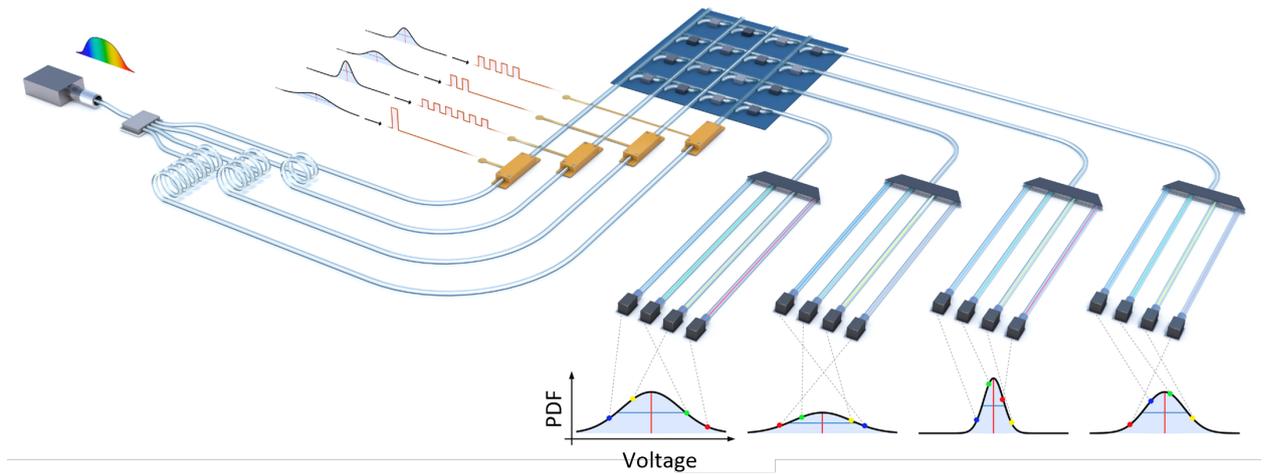

**Fig. 1. Photonic probabilistic processor.** We generate broadband chaotic light via amplified spontaneous emission. By splitting and delaying the light beyond the coherence time of the source, we create four uncorrelated optical carrier signals. On each signal arm, we encode an input waveform which determines the noise distribution. A photonic crossbar array based on the non-volatile phase change material Germanium-Antimony-Telluride performs multiply and accumulate operations on the input states. At the output we demultiplex the broadband optical signal to independently sample from the output distributions in parallel.

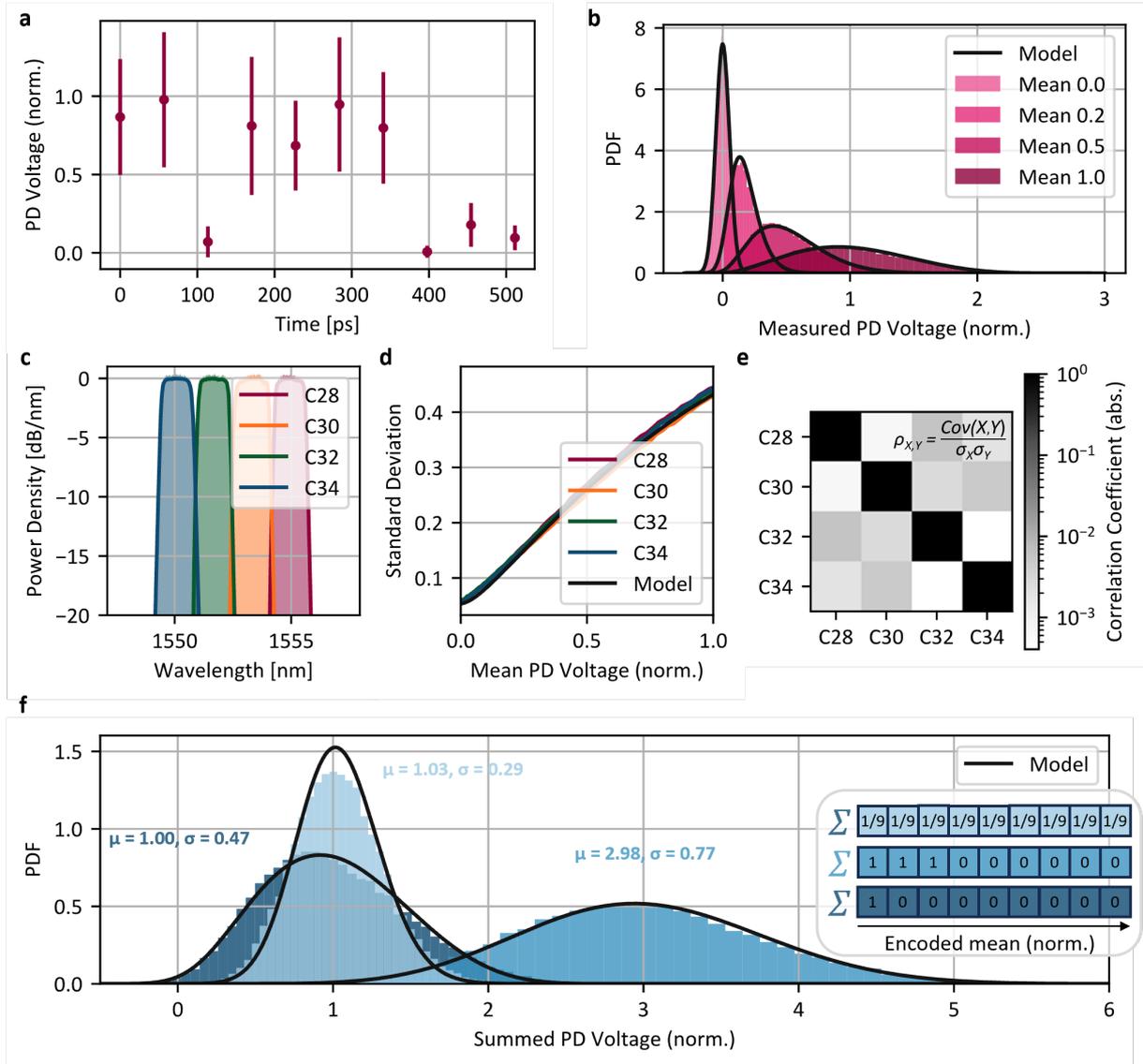

**Fig. 2. Chaotic light as an entropy source. a,** We encode the distributions with a symbol length of 56.8ps on the chaotic optical carrier. Due to the intensity fluctuations of the ASE, the standard deviation of the measured signal increases with the mean, as indicated by the error bars. **b,** For small intensities and hence low voltages at the photodetector, the measured probability density function (PDF) is dominated by the electronic noise of the readout system. For increasing mean voltages, the optical intensity fluctuations are the dominant source of randomness, and the distribution becomes Bose-Einstein like. **c,** Spectral output of the broadband chaotic light at four channels of the 200 GHz ITU grid for parallel sampling. **d,** For each wavelength channel, the relation between standard deviation and mean of the distribution is shown. For all channels there is ground noise due to electronic noise and a limited extinction ratio of the modulators, afterwards the standard deviation increases linearly due to the chaotic carrier until detector saturation limits the noise. **e,** Measured correlation

coefficients between the detector voltages in the different wavelength channels during parallel sampling. The correlation between different channels is below $10^{-2}$, which enables independent sampling. **f,** Shaping the form of the distribution by taking the sum of 9 subsequent symbols. The mean of the distribution is the sum of the means, whereas the variance depends on the actual waveform.

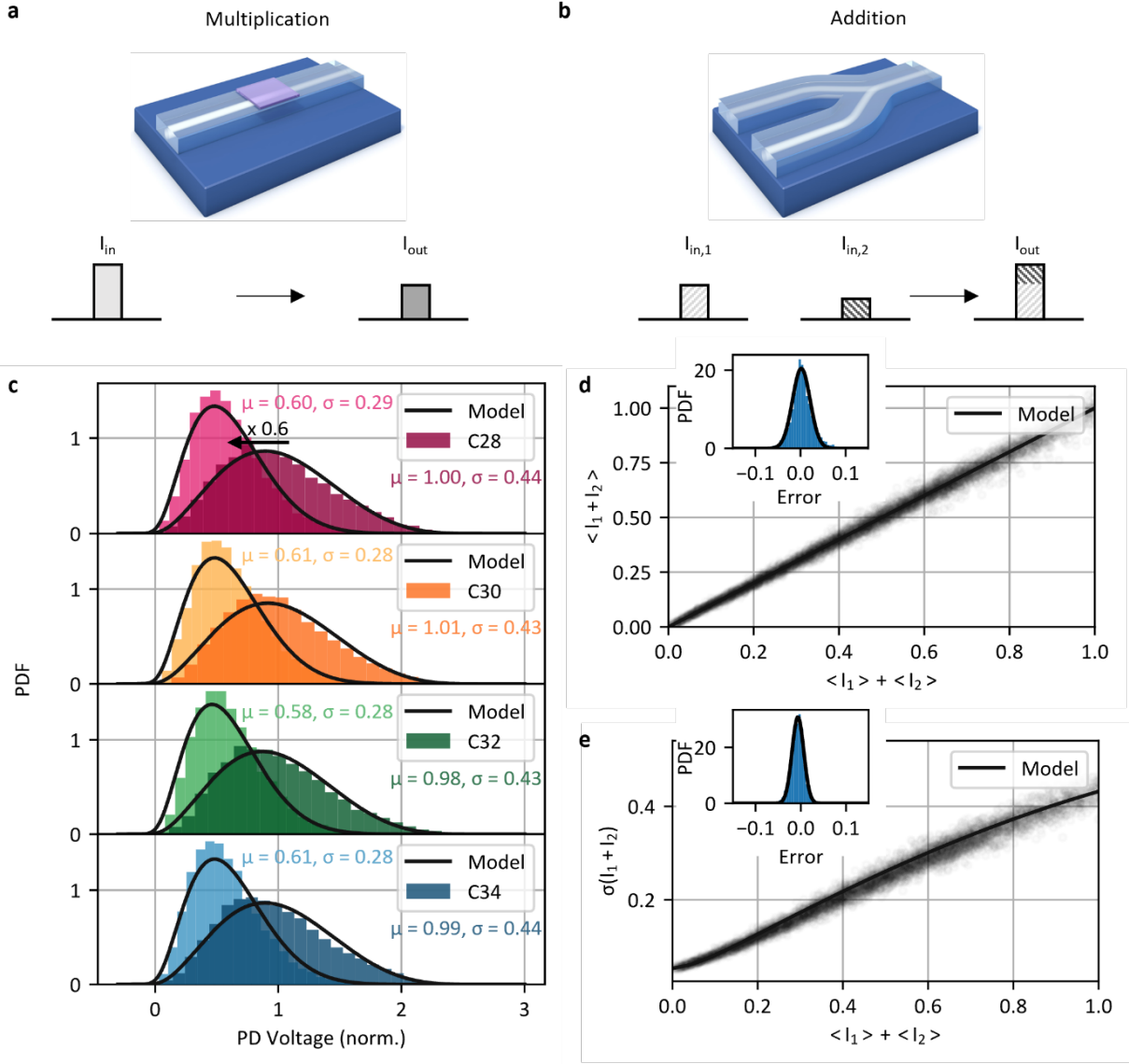

**Fig. 3. Arithmetic Operations. a,** Signal information is encoded in the intensity of the chaotic light field. Non-volatile tunable GST attenuators are used as the multiplication and memory units. **b,** Two input distributions are summed up by overlapping them in a single waveguide. Due to the random spectral phase of the chaotic carrier the means of the input distributions add up. **c,** Sampling from the output distribution in four wavelength channels in parallel. We set a relative transmission of 0.6 in the GST cell and choose an input distribution with mean 1 for all channels. **d-e,** Comparison of the mean and standard deviation of the output distribution with the properties of the input distribution during parallel sampling. As expected, the mean of the output distribution is the sum of input means. The measured mean aligns with the model prediction within 0.4 % and has a spread of 2.6 %. Similarly, the standard deviation of the output distribution matches the model prediction with an average deviation of 0.6 % and a spread of 1.2%.

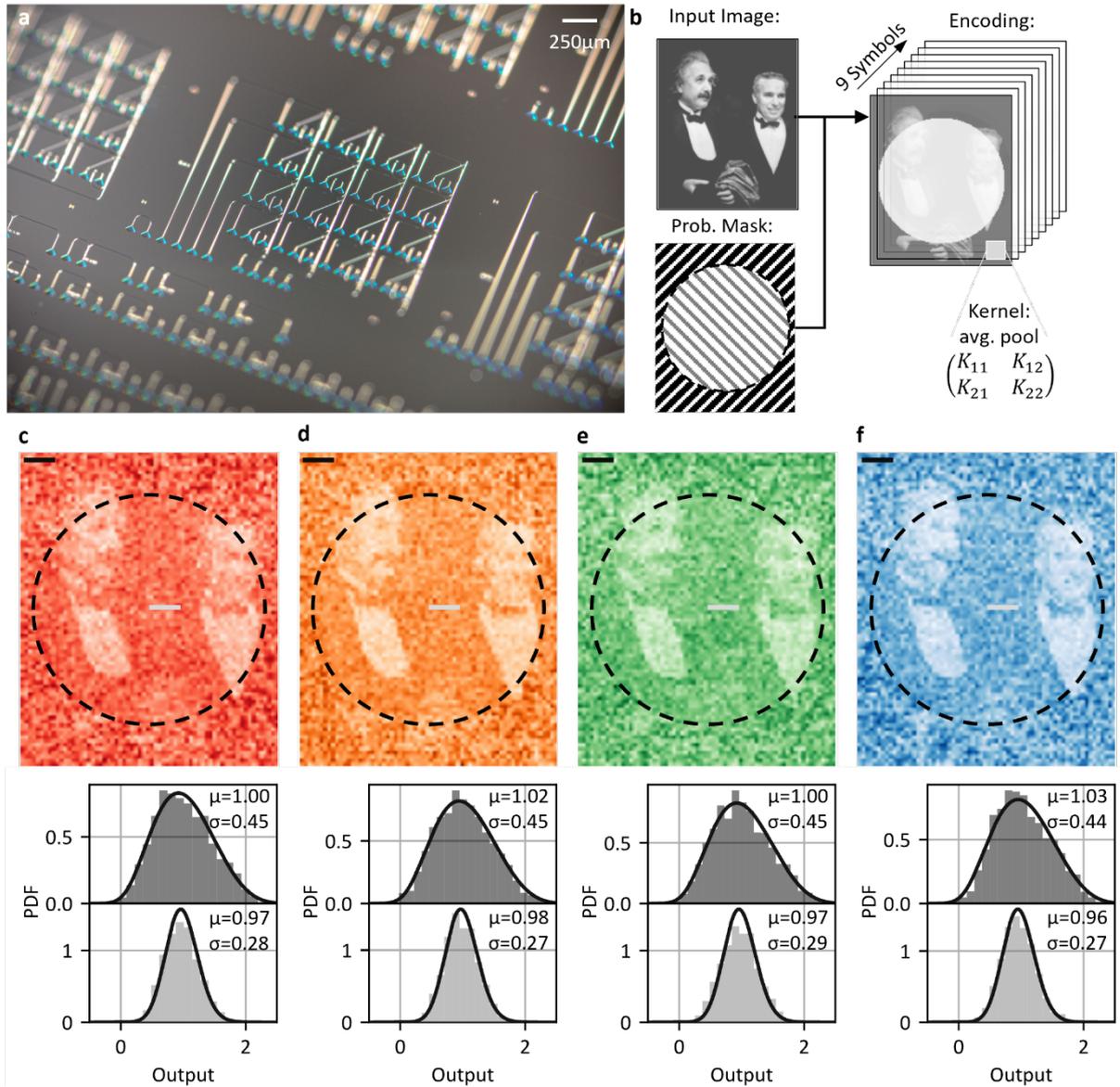

**Fig. 4. Probabilistic convolution processing. a,** Optical microscope image of a 4x4 photonic crossbar array and calibration structures. **b,** We convert the deterministic input image pixel wise to probability distribution via the encoding shown in Fig. 2 sketched by the probabilistic mask. For all distributions the mean corresponds to the pixel value. In the inner region we reduce the variance by distributing the mean over all nine symbols whereas we encode the mean in a single symbol in the outer area. Then we perform the convolution operation by sliding the kernel over the input image. **c-f,** Parallel sampling on four WDM channels for average pooling with stride two. The probabilistic mask leads to a larger standard deviation in the outer area of the picture than in the inner area.

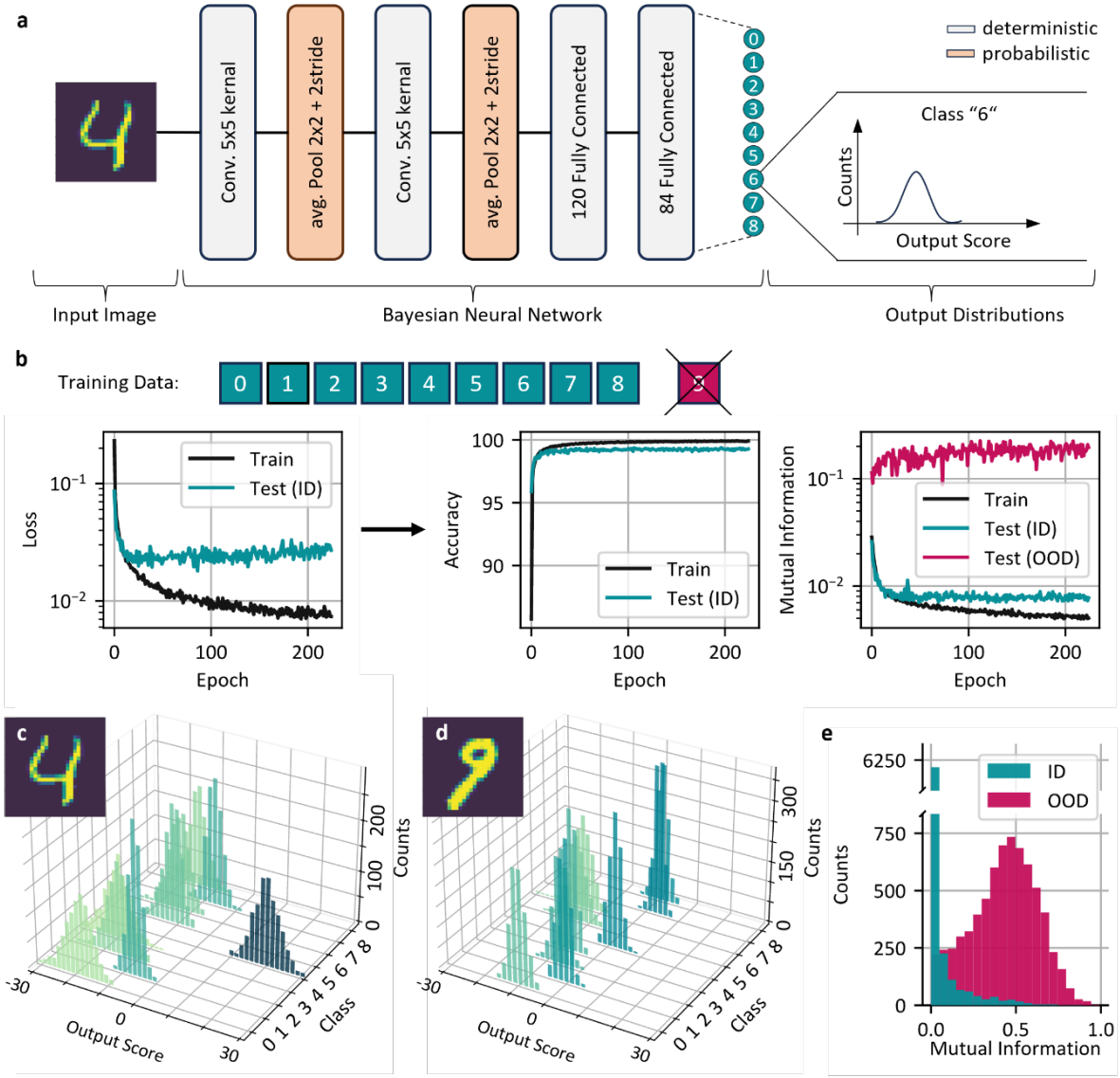

**Fig. 5. Bayesian inference on the incomplete MNIST dataset. a,** We employ a modified LeNet-5 model for digit classification on a "nine-class MNIST" dataset. Average pooling (2x2 kernel) operations are performed with the photonic crossbar array which enables probabilistic modelling. For a single input image, the BNN predicts the distribution of each output class. **b,** The Bayesian model is trained via stochastic variational inference. During the training, the loss (inverse evidence lower bound) is minimized resulting in a maximized classification accuracy and a stark contrast in the average Mutual Information of the output distribution between in-domain (ID) and out-of-domain (OOD) images. **c-d,** We sample from the BNNs output distribution for an ID image and an OOD image. For the ID image, the network clearly assigns the highest activations to the correct class label whereas the distributions overlap for the OOD image. **e,** Evaluating the BNN for all ID and OOD test images shows a stark contrast between the mutual information, successfully rejecting the OOD ones.

24